# Blockchain for IOT-based NANs and HANs in Smart Grid

Shravan Garlapati, Email: gshra09@vt.edu

## ABSTRACT

Smart Grid – an intelligent connected grid consisting of millions of smart devices, used to collect data from the grid to improve the efficiency of its operation. These smart devices communicating wirelessly are susceptible to attacks and hence the security and privacy of the smart devices along with the smart grid is a major challenge. Blockchain-based systems provide improved security and privacy and hence gained a lot of attention in the recent past. This paper proposes a blockchain-based architecture for Neighborhood Area Networks (NANs) and Home Area networks (HANs) in Smart Grid. The paper presents a security analysis in terms of confidentiality, integrity and availability to show that the proposed blockchain-based Smart Grid architecture is secure. Also, the impact of the improved security on packet delays, energy and computational overhead is discussed.

## INTRODUCTION

In many advanced countries, the aged power grid is being renovated to be an intelligent power grid to improve the efficiency of power generation, transmission and distribution. In this process, the power grid is being installed with smart devices and sensors to automate and improve the efficiency of different applications such as Advanced Metering Infrastructure (AMI), Advanced Distribution Automation (ADA), Demand Response Management System (DRMS) and Outage Management System (OMS). In order to communicate with the smart devices, Smart Grid (SG) generally employs a two-way communication network. In the Distribution Smart Grid (DSG), the two-way communication network includes multiple sub-networks such as Neighborhood Area Networks (NANs) and Home Area Networks (HANs). A HAN manages the communication between smart devices within a home. A NAN is used for two-way communication between the neighborhood data aggregation point and HANs. A Home area Gate Way (HGW) and Neighborhood Gate Way (NGW) manages the inbound and out-bound traffic of HAN and NAN respectively. It is well known that the SG can employ different types of networking technologies like Wireless Mesh Networks (WMNs) and long-range single-hop wireless networks. According to [1] and [2], due to its multi-hop structure and to ensure 100% connectivity of devices in SG, WMNs is favored over single-hop wireless networks like cellular. But this work assumes that both the cellular networks and WMNs are employed in DSG (Figure 1) to take advantage of the faster single-hop communication and 100% connectivity offered by WMNs.

    The DSG applications i.e. AMI, ADA, DRMS and OMS consists of smart devices and sensors that generate, process and exchange huge amounts of secure, safety-critical and privacy-sensitive data and hence the SG can be a target of cyber-attacks. An adversary gaining access to a single device has the potential to disrupt the entire power grid of a nation. The security and privacy requirements of SG networks are different when compared to IT networks. The security in IT networks is more focused on providing protection at the network center (where the data is stored), whereas the protection in SG networks is needed at the network center and edge [3] i.e. *distributed* in nature. Numerous studies have proposed architectures to improve the security and privacy of SG using the traditional state-of-the-art security solutions such as IDS, firewalls and encryption methods. However, these conventional mechanisms have limitations (for example: highly centralized access control) and hence they are unsuitable for SG [4][5]. Moreover, scalability is an important factor as a utility may install millions of

smart devices/sensors in the field. Hence, SG communication networks demand a scalable and distributed secure and privacy frameworks. Blockchain (BC) technology - the building block of the cryptocurrency systems like Bitcoin, Ethereum, ripple etc., has the potential to overcome the above-mentioned challenges due to its distributed, secure and private nature.

BC can be defined as a specific type of distributed ledger that records transactions between two parties in an efficient and secure manner. In simple terms, BC is a growing list of records called as *blocks*, which are linked using cryptography [6]. A block contains transaction data, cryptographic hash of the previous block and a timestamp. As the blocks are connected using cryptographic hash, a block in BC cannot be easily altered and hence the data is resistant to modification. In order to operate as a distributed ledger, a BC based system is typically managed by a peer-to-peer network. In BC based systems like Bitcoin, each user is identified by a changeable public key. In bitcoin network (BCN), certain users with large computational resources are categorized as *miners*. In order to transfer money, a user generates and broadcasts a new transaction to the network. Every new transaction in a BCN is validated by the miners and a valid transaction is pushed on to the block. A block in a BCN consists of multiple transactions and whenever a block is full, it is appended to the BC by a mining process [7]. As a part of the mining process, miners solve a cryptographic puzzle known as *Proof of Work (PoW)*. A miner node that first solves the *PoW* appends the mined block to the BC and broadcasts the solution to the network. All the other miners in the network validate the solution, accept the updated BC and re-broadcast the new block. The miner node that solved the *PoW* is rewarded with bitcoins.

Adopting BC technology to the SG comes with several challenges such as: demand for high computational resource to solve *PoW*, long latency to validate the transactions, resulting in low transaction rate (bitcoin - 7 transactions/sec, Ripple – 15 transactions/sec) and scalability issues due to broadcasting transactions and blocks to the entire network. Ali Dorri *et al* proposed a novel BC for Internet of Things (IoT) by eliminating the need for PoW and bitcoins [8]. Using Smart Home as a case study, they illustrated the use of BC to improve the security and privacy for IoT [9]. In this paper, with some modifications, we adopt the ideas presented in [8] and [9] and propose a BC-based system to improve the security and privacy of IoT based HANs and NANs in the DSG. The proposed architecture is as shown in Figure 1 (details discussed later). A local BC (LBC) is employed to provide secured access control to the smart devices in HANs and NANs. Also, both the LBC and a BC in remote storage maintain a chronologically ordered history of immutable transactions of smart devices in DSG. In this study, based on the qualitative analysis, we demonstrate that the proposed architecture achieves *confidentiality*, *integrity* and *availability* and discuss its potential to thwart key security attacks such as linking, Distributed Denial of Service (DDoS) and modification attacks. The main contributions of this paper are:

i) In [8] and [9], the transactions are classified into five different categories: genesis, store, monitor, access and remove transactions. Based on the DSG application needs, this paper introduces new transaction types such as: *event-based transaction (EBT)*, *OTF-read transaction (OTFT)*, *control transaction (CT), alarm transaction (AT)*.

ii) Proposed a BC-based architecture for DSG as shown in Figure 1. Generally, in cryptocurrency BC systems, every node has a copy of the BC i.e. if the BC network has *N* nodes, there are *N* copies of BC. But in the proposed architecture, there are only two copies of BC for each device i.e. LBC and remote BC (RBC), *distributed* at the network edge and central storage. In the BC-based smart home presented in [9], there is only one copy of data in LBC. Maintaining a single copy of data is a risk as data can be lost in the case of an attack or hardware failure. Hence, the proposed architecture offers redundancy by storing data in both LBC and RBC.

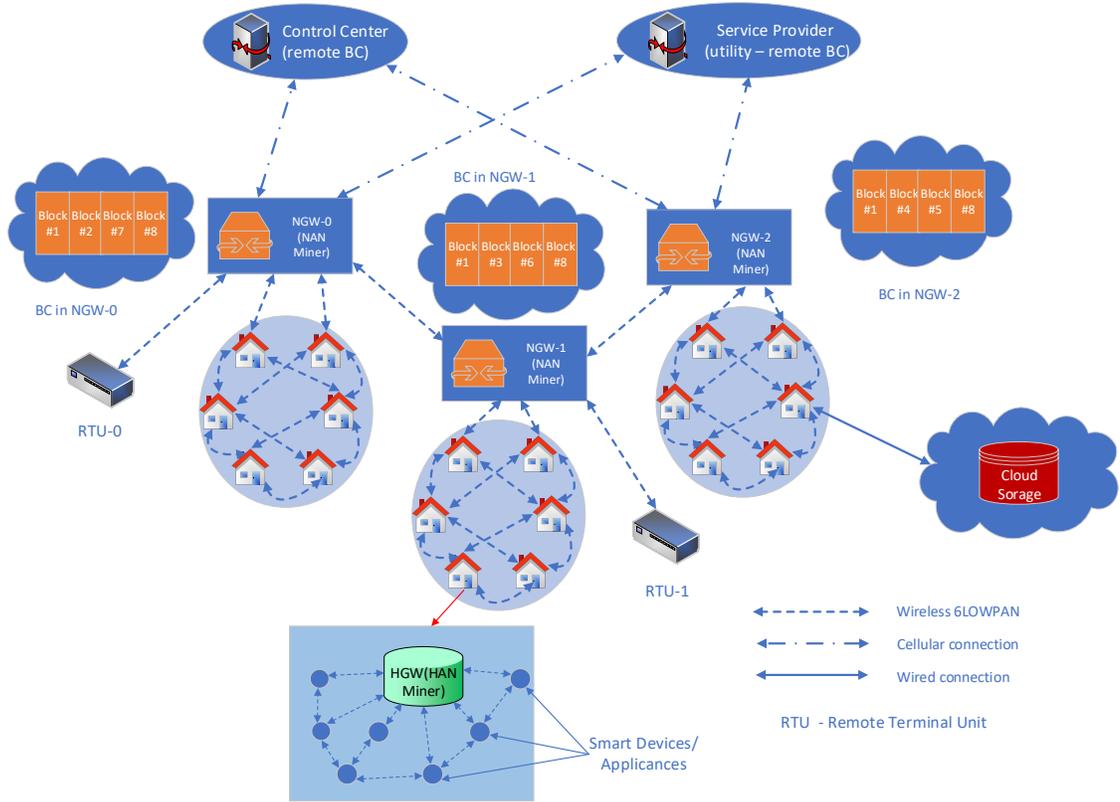

Figure 1 BC-based Distribution Smart Grid

iii) Provided qualitative analysis on the security and privacy of BC-based DSG against DDoS, linking and modification attacks. Proposed a novel way of exploiting the multicast/broadcast transactions to improve the security against modification attacks in DSG.

Remainder of the paper is organized as follows: First, the core components of the BC-based architecture are presented. Next, the architecture of the BC-based DSG is discussed in detail. Subsequently, security and BC-overhead analysis are presented before concluding the paper.

## MAIN COMPONENTS

The details of the core components of the BC-based architecture presented in Figure 1 are discussed in this section.

A. **Transactions:** In BC-based DSG, any exchange of data between devices in HAN and NAN are known as *transactions*. It is assumed that there exist different transactions in BC-based DSG, each designed for specific purpose. In addition to the genesis, store, monitor, access and remove transactions presented in [9], this paper introduces new transaction types such as: EBT, *OTFT*, *CT*, AT. The purpose and the usage of these transactions is explained as and when needed in the next sections. In order to secure the communication, it is assumed that all the transactions use a shared key. Also, it is assumed that a lightweight hashing is used to detect any changes in the content of the transaction during transmission [9] [10].

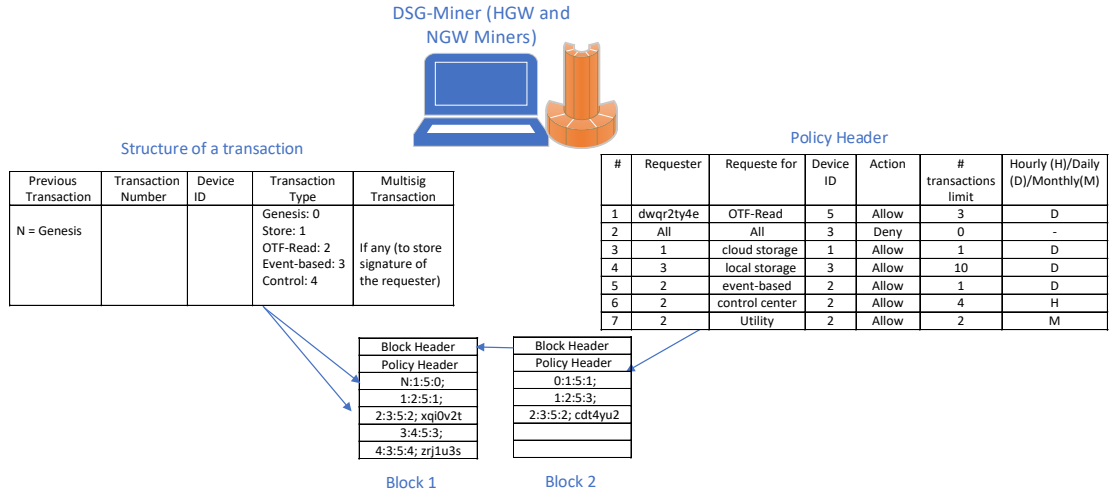

Figure 2 DSG-BC including the structure of a transaction and the policy header

B.  **DSG-BC:** In DSG, a BC keeps track of inbound and outbound transactions, authorizes devices and based on a policy header enforces policies. Similar to cryptocurrency BCs, considering *genesis transaction* as the first transaction, all the transactions of a device are cryptographically chained together as an immutable ledger in DSG-BC. In order to provide immutability and enforce policies, a block in DSG-BC is designed with two headers, they are: block header and policy header. The block header contains the hash of the previous block to keep the DSG-BC immutable. The policy header is used to authorize devices and enforce policies on transactions. The structure of the policy header and the transaction are as shown in Figure 2. The policy header structure is similar to the structure presented in [9] and unlike four parameters in [9], it has six parameters. They are: *Requester* (identified with Public Key), *Request Type* (transaction type - *store, access, control* etc.), *Device ID* (SM ID or thermostat ID etc.), *Action* (Allow or Deny the request), *number of transactions limit* and *limit duration* (hourly, monthly and daily). Apart from the headers, a block contains several transactions. In BC, for each transaction upto five parameters are stored. The parameters *Previous Transaction* and *Transaction Number* aid in uniquely identifying a transaction and chaining transactions of the same device. The third parameter *Device ID* indicates the ID of the device from which the transaction originates. The fourth parameter is the *transaction type* (EBT, OTFT, CT etc.). The fifth parameter identifies a *multisig transaction* and stores the signature of the requester. In this study, as mentioned above, it is assumed that there exist two types of BCs. They are: LBC and RBC. Depending on its location, LBC is categorized as HAN-BC and NAN-BC. A HAN-BC is maintained by the HGW miner whereas a NAN-BC is managed by the NGW miner. As a SM sends data to both the CC and utility storage, there exists an RBC at both the CC and utility, respectively known as CC-RBC and U-RBC. As the purpose of the U-RBC is to store/manage the data, *store* transaction (ST) and *read* transaction are the only allowed transaction types in its policy header. On the other hand, the CC-RBC handles *OTFT*, *CT, AT* and *EBT* along with the *store* and *read* transactions. Hence, all the transaction types are allowed in its policy header.

C.  **DSG-Miner:** A DSG-Miner handles several key functions such as authentication, authorization, auditing transactions, generating genesis transactions, distributing and updating keys, maintaining the transaction structure. Also, the miner processes inbound and outbound transactions in its network (HAN or NAN). The miner collects transactions into a block and once a block is full, appends it to the BC. The DSG-Miner can be built into the internet gateway or a standalone device that can be placed between gateway and devices. As mentioned above, the proposed architecture has three different types of miners, they are *HGW miner*, *NGW miner* and *storage miner (RBC)*. The number of devices handled by HGW can be around 10, NGW can be around 100 and storage miner are in the order of 1000s. HGW and NGW miners maintain 2 copies of BCs i.e. control center BC and utility BC for each SM handled by them. Figure 3 shows the arrangement of data in LBC (HGW, NGW) and RBC (CC-RBC and U-RBC). As shown in Figure 3, for each SM, two copies of data are stored, one in LBC and other in RBC.

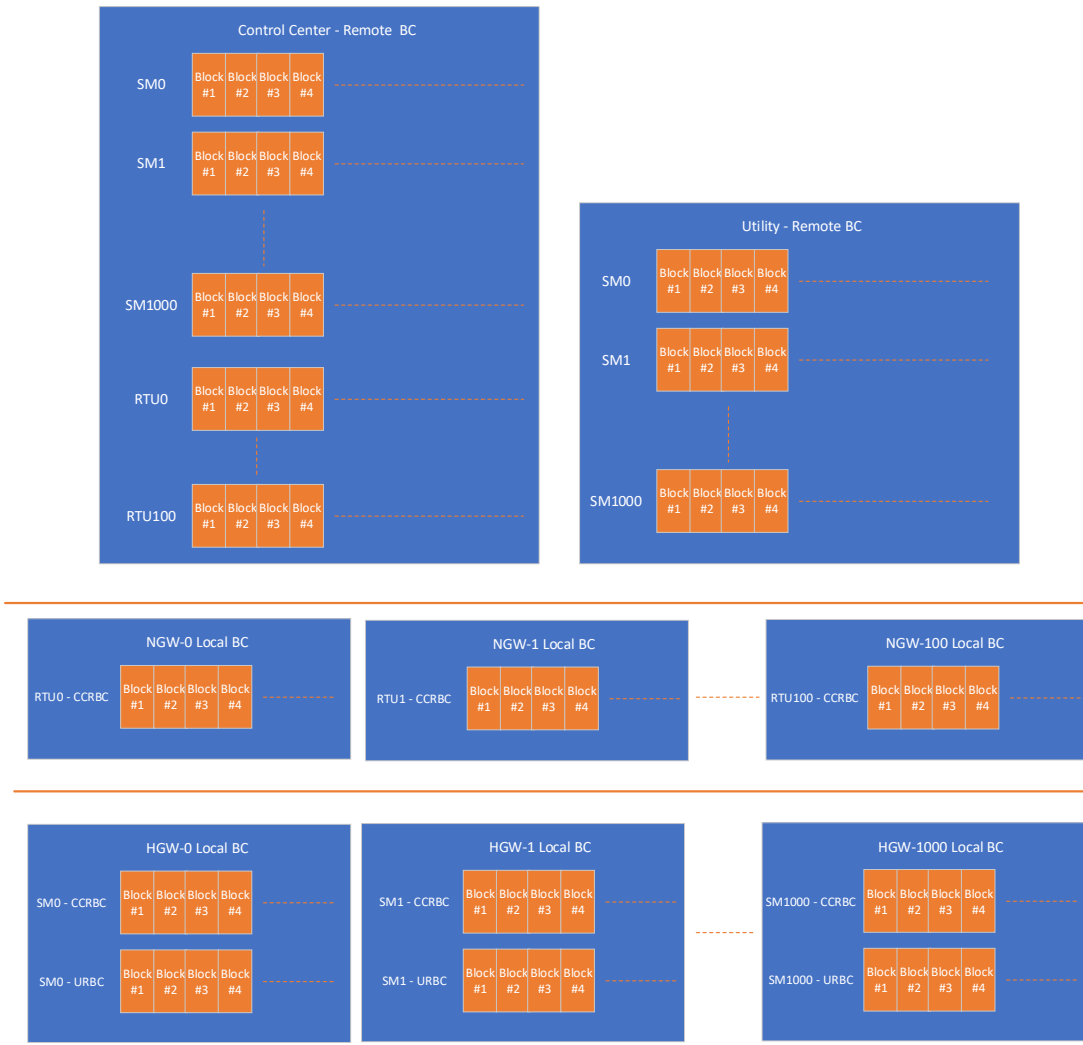

Figure 3 Arrangement of data in Local BC and Remote BC

D. **Overlay Network:** As shown in Figure 1, HGW and NGW miners constitute an overlay network along with cloud storage, utilities, CC and remote terminal units (RTUs). The overlay network is similar to the peer-to-peer network in bitcoin and offers *distributed* feature to the proposed architecture. According to [2] and [9], as shown in Figure 1, 6LoWPAN (*IPV6 over Low-Power Wireless Personal Area Networks*) technology is used for NAN and HAN communications. It is assumed that the external communication (i.e. to utility or CC) of a NAN is based on long-range cellular. Also, wired public IP network is assumed to be used to store data to cloud.

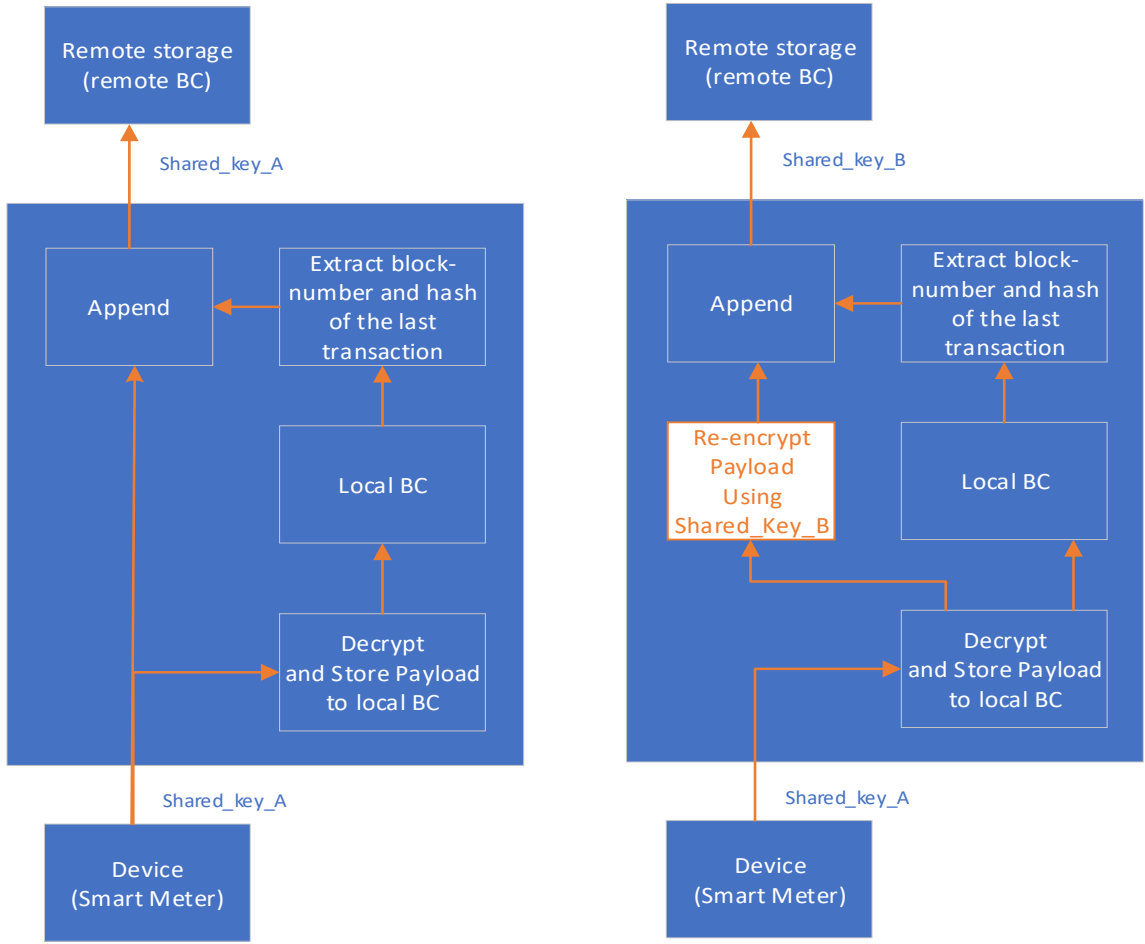

Figure 4 Illustration of data flow between device and remote storage using single shared key and two shared keys

**THE BC-BASED DISTRIBUTION SMART GRID**

This section describes the initialization steps of adding new devices, defining policy header and transaction handling in the BC-based DSG.

A. **Initial Setup:** In order to add a new device to the HAN or NAN, the corresponding DSG-miner creates a *genesis transaction* by sharing a key with the device using generalized Diffie-Hellman (DH) [11]. In the case of SMs, as the data is also stored to RBC, key must be shared with the storage miner as well. As the key must be known to 3 entities i.e. device, storage miner and local miner, the following two different mechanisms can be employed to exchange the key:
   a. Two different DH key exchanges i.e. one between local miner and the device and the other between the local and remote miners are done. Hence, different shared key is used to exchange data between device and local miner and between local and remote miners.

  b. Single DH key exchange between 3 entities. In this case, same shared key is used for data exchange between the device and local miner and the local and remote miner.

Figure 4 illustrates the ST data flow based on the above mentioned two schemes. The difference between two schemes is self-explanatory. The shared key is stored in the genesis transaction of the device's LBC and RBC. This work assumes that each device has its own BC. In the case of SMs, LBC i.e. HAN-BC maintains two BCs per each SM, corresponding to CC-RBC and U-RBC. Regarding the policies, using a software interface provided by the company that installs the HAN-BC, home owner creates/updates policies for HAN based on the policy header structure presented in Figure 2 and adds it to the first block of its HAN-BC. A portion of the policies in HAN are setup/updated exclusively by the external entities such as utilities to create/update policies related to SMs (electric, gas and water). Home owners are not authorized to update the policies related to SMs as they may modify the settings to reduce the utility charges. The policy header in the NAN will be initially setup by the utility and any future updates will be done via internet. A DSG-miner makes decisions based on the policy header from the most recent block in BC. Hence, policy updates are always applied to the current block's policy header.

B. **HAN Transactions:** A device in HAN may operate based on the data requested from another device. For example, Thermostat controls AC (turn on/off) based on the data collected from the temperature sensor. In order to secure the data exchange and keep track of the transactions, a shared key is allocated (based on the policy header) by the HGW miner to devices that exchange data. Devices can exchange data as long as the key is valid. The miner invalidates the key by sending control messages to the devices. A device can store data to LBC and/or RBC using a ST. In [8] and [9], procedure to store the data to cloud i.e. remote storage is discussed. It employs anonymous authentication but the mechanism to secure the data transfer is not explained. This section discusses the mechanism to securely transfer the data to remote storage.

  A SM generates a ST to store data to both LBC and RBC. When an HGW miner receives a ST from an SM, as shown in Figure 4, it decrypts and stores the payload to the HAN-BC. Next, the miner extracts the hash and the block-number of the last transaction from the HAN-BC, appends it to the new data, creates a new ST and sends it to the RBC. If the block-number and the hash of the last transaction extracted from the LBC match with ones stored in RBC, SM is authenticated. As the RBC has the shared key, it decrypts the data after authentication. At the end of a ST processing, RBC and LBC of a device are synchronized w.r.t blocks, transactions in blocks and hash of the data. The same procedure can be applied to any HAN device that intends to store data securely to the cloud.

C. **NAN Transactions:** Similar to HAN, the data exchange between NGW miner and an RTU in a NAN is based on a shared key allocated by the NGW miner. The mechanism to process *non-HGW miner* (RTU) store transactions is similar to the HAN store transactions discussed above. In the case of HGW miner (ex: SM transactions), NGW miner authenticates the transaction and forwards it to the appropriate destination.

D. **DSG Transactions:** Most of the DSG transactions are hierarchical transactions i.e. the data exchange between the utility or CC and a HAN happens via a NAN. An *OTFT* is generated by CC to read the power consumption data, on-the-fly from the SM. When the NGW miner receives a read request from the CC, the miner validates the request and forwards the request to the corresponding HGW miner. HGW miner authenticates the request, reads data on-the-fly from the SM and sends it back to the CC via NAN.

An *EBT* is generated by a SM to report an event such as power outage. The data encapsulated in an *EBT* is sent to the CC and the data transfer mechanism is similar to the ST described earlier. In DRMS, CC remotely controls the smart electrical devices in a customer's home depending on the peak electricity demand. In order achieve this, a *CT* is sent by the CC to turn on/off a smart device. The data exchange mechanism of a *CT* between the CC and a smart electrical device in a HAN occurs via NAN and it is similar to an *OTFT*.

## SECURITY ANALYSIS

This section analyzes the security and privacy of the BC-based DSG. In order to maintain the security of information in SG and keep it protected, the National Institute of Standard and Technology (NIST) has defined three main requirements, namely: **Confidentiality, Integrity and Availability,** known as CIA [12]. Confidentiality ensures that only the authorized entity can access the information. *Integrity* ensures that the transmitted data is received at the destination without any modifications and *availability* makes sure that the computational and communication resources are available to a device when it is needed. Table 1 summarizes the mechanisms employed by the proposed BC-based architecture that aid in resisting against device, data and privacy attacks to meet the security requirements in DSG.

Table 1 Security Requirements Analysis

| Requirement | Mechanism Employed |
|---|---|
| Confidentiality | Symmetric encryption is employed to achieve confidentiality |
| Integrity | Achieved using hashing |
| Availability | Achieved by limiting the number of valid transactions per device and the miner |
| User Control | Achieved by transaction logging in local BC |
| Authorization | Policy header, block hash and shared keys are used to authorize devices |

To increase the network resource availability, smart devices should be protected from malicious requests. As per [9], this requirement can be met by limiting the number of transactions (per hour or per day) to those entities with which each device has established a shared key. In the proposed architecture, as shown in Figure 2, the limit on the number of transactions is stored in the policy header. As all the transactions are authorized by the miners, the limit will be enforced by them. There are two types of AMI data: low-frequency data (once per week) sent to utilities for billing purpose and high-frequency data (3-4 times per hour) contains power usage patterns sent to CC, used for real-time control and optimization [5]. Hence, the limit on the number of STs for low-frequency and high-frequency data is once per week and 3-4 times per hour respectively. In DRMS, utility can send control messages to turn on/off customer's AC multiple times per day. Hence, the limit on the DRMS CT is 3-4 times per day. Thus, the limit on the number of transactions can be varied depending on the type of the data and type of the DSG application.

DDoS attack is a major attack that affects the *network availability* [5]. In the literature, multiple studies analyzed the impact of DDoS attack on smart grids [13]. Also, the SM data can be analyzed to establish the link between electricity consumption and customer presence in the home to perform physical attacks such as robbery. Also, as the DSG collects and stores the data, it is important to study the data modification attacks. Hence, we analyze the effectiveness of the proposed solution to handle the critical security attacks such DDoS, linking and modification attacks in DSG.

A. **DDoS Attack:** The proposed design offers multiple levels of defense against this attack. As the SG devices are directly inaccessible, the first level of defense comes from the fact that it is extremely difficult for an adversary to install malware on them. Despite this inaccessibility, if the adversary somehow manages to obtain access to the device and attempts to flood the DSG network, the miner offers the second level of defense as it monitors all the outbound traffic. As the policy header puts a limit on the number of transactions, if a compromised device generates an abnormal amount of traffic, the miner will send an *alarm transaction* to the concerned authority (sending a message to the customer in the case of HAN and utility in the case of NAN) about the excessive traffic. The concerned authority takes the necessary steps to identify the compromised device and recover from the attack. During the identification and recovery process, the compromised device may continue to generate the abnormal traffic. Hence, the proposed design assumes that the miner is equipped with multiple communication and computational resources such that the legitimate traffic flows uninterrupted even during attack, compromised device identification and recovery. Also, the miner does not forward the abnormal traffic into the network. The limit on the number of transactions is applied to the outbound transactions of miners as well.

B. **Linking Attack:** To protect against this attack, each SG device's data is shared and stored using a unique key. Also, the key can be varied depending on the type of the data. For example: as mentioned before, SM sends two types of messages, low-frequency data destined to the utility and high-frequency data to the control center. So, to increase the protection level, two different keys i.e. Low Frequency Unique Key (LFUK) to store low-frequency data and High Frequency Unique Key (HFUK) to store high-frequency data. For each devices data type, DSG-BC creates unique ledger of data using different PK.

C. **Modification Attack:** In order to launch this attack, the attacker must compromise the storage security. The adversary may then attempt to delete or modify the data of a device. As mentioned above, in the proposed architecture, a device generating the ST is authorized by storage miner by comparing the block number and hash of the last transaction of the device against the ones stored in the RBC. Hence, any change in the device's stored data will be automatically detected during a new *store transaction*. In order to complicate the scenario, let's assume that along with the storage, HAN-0 device's LBC is also compromised and hence the attacker can modify the device's data in LBC and RBC. Hence, the comparison of device's hash at the RBC and LBC cannot detect the attack. In this case, we propose the idea of exploiting the multicast/broadcast packet transmissions in SG networks to aid in detecting the data breaches. According to [14], electricity pricing and DRMS control data is communicated as multicast/broadcast transactions to the SMs by the utility/control center. As shown in Figure 1, broadcast transactions 1 and 8 are part of the LBCs at NGWs, HGWs and storage. Hence, NGW-1 and NGW-2 can generate read transactions (periodically) to read HAN-0 device's broadcast data from the storage and independently compare with their own LBC broadcast data. If HAN-0 broadcast data mis-compares with their own data, NGW-1 and NGW-2 will send an *alarm transaction* to the control center indicating that the HAN-0 device is compromised. In a larger SG network, the number of NGW's can be in 100s. Hence, for each

HAN device that stores data to storage, we must select the corresponding NGWs to periodically read and compare the broadcast transactions. The procedure to select the NGWs, how many NGWs must be selected for reading/comparing and how often broadcast/multicast packets are compared are part of the future study.

**BLOCK CHAIN OVERHEAD ANALYSIS**

As discussed above, BC-based system has the potential to improve the security and privacy of a DSG. But the improved security and privacy can result in packet, energy and computational overhead. According to the simulations presented in [9] for BC-based smart home, the increased payload sizes due to encryption and hashing in BC-based systems have relatively smaller effect compared to the lower layer header overhead in 6LoWPAN. Also, delay overhead is ~20 msec and the energy consumption of the 6LoWPAN devices increases by ~60% [9]. Unlike the transmission SG [15], where there are stringent timing requirements (< 1 sec), DSG applications are not time critical. Hence, 10's of msec of delay overhead is not an issue. Unlike electricity SMs, that rely on power from grid, gas and water SMs are battery operated. Therefore, for gas and water SMs, it is a trade-off between the energy overhead (~60%) and the improved security offered by the BC-based systems. As the traffic generated by the wireless SMs is not very high, they operate for many years without replacing the battery. Hence, our recommendation is that 60% overhead is not very high, and the security and privacy of BC-based systems outweigh the energy overhead.

As mentioned before, HGW/NGW miners monitor the inbound and outbound transactions, enforce the policy header rules, manage shared key for data exchange and maintain the LBC. These operations incur energy and computational overhead. As discussed above, the proposed architecture assumes that miner is integrated as a part of internet gateway and unlike the smart devices that are battery operated, miner has a continuous power supply. Hence, energy consumption is not an issue. As mentioned above, SMs generate 3-4 packets per hour, DRMS generates 3-4 packets a day, during an outage OMS generates 1 packet. Hence, the average number of packets generated by DSG applications is low i.e. around 10-15 per hour. Therefore, the computational and storage resources needed to maintain LBC for DSG applications is not very high. If home automation devices (ex: google NEST), are also included in HAN-BC, the storage and computational resource requirement may increase.

**CONCLUSION**

In summary, this paper presented a Block Chain based architecture for NANs and HANs in distribution smart grid. Also, discussed various components, transactions and processes involved in the data exchange. Unlike the traditional centralized security systems, the proposed BC-based system distributes the security at network edge and center. The paper presented a security analysis to qualitatively explain the effectiveness of the proposed architecture to handle security attacks such as DDoS, linking and modification attacks. In this process, a novel way of using broadcast/multicast transactions to detect data modification attacks is proposed. Also, based on the BC overhead analysis, it is concluded that the delay, energy and computational overhead is not significant and hence is not an impediment to build the BC-based Distribution Smart Grid.

# CITATIONS

1) Neighborhood Area Communication Network - https://www.nist.gov/programs-projects/neighborhood-area-communication-network
2) Dong Chen, J. Brown and J. Y. Khan, "6LoWPAN based Neighborhood Area Network for a smart grid communication infrastructure," *2013 Fifth International Conference on Ubiquitous and Future Networks (ICUFN)*, Da Nang, 2013, pp. 576-581.
3) Fadi Aloul, A.R. Al-Ali, Rami Al-Dalky, Mamoun Al-Mardini, Wassim El-Hajj "Smart Grid Security: Threats, Vulnerabilities and Solutions" *International Journal of Smart Grid and Clean Energy*, Vol. 1, No. 1, September 2012: pp. 1-6.
4) Zakaria Elmrabet, Naima Kaabouch, Hassan El Ghazi, Hamid El Ghazi "Cyber-security in smart grid: Survey and Challenges" *International Journal of Computers & Electrical Engineering,* Vol 67, April 2018, Pages 469-482.
5) X. Li, X. Liang, R. Lu, X. Shen, X. Lin and H. Zhu, "Securing smart grid: cyber-attacks, countermeasures, and challenges," in *IEEE Communications Magazine*, vol. 50, no. 8, pp. 38-45, August 2012.
6) https://en.wikipedia.org/wiki/Blockchain
7) S. Nakamoto, "Bitcoin: A peer-to-peer electronic cash system," 2008.
8) Ali Dorri, Salil S. Kanhere, Raja Jurdak "Blockchain in Internet of Things: Challenges and Solutions"
9) A. Dorri, S. S. Kanhere, R. Jurdak and P. Gauravaram, "Blockchain for IoT security and privacy: The case study of a smart home," *2017 IEEE International Conference on Pervasive Computing and Communications Workshops (PerCom Workshops)*, Kona, HI, 2017, pp. 618-623
10) A. Bogdanov, M. Knezevic, G. Leander, D. Toz, K. Varici, and I. Verbauwhede, spongent: A Lightweight Hash Function. Berlin, Heidelber: Springer Berlin Heidelberg, 2011, pp. 312-325.
11) H. Delfs, H. Knebl, and H. Knebl, *Introduction to cryptography*. Springer, 2002, vol. 2.
12) S. G. I. Panel, "Guidelines for smart grid cyber security: Vol. 1, *smart grid cyber security strategy, architecture, and high-level requirements*, and Vol. 2, privacy and the smart grid, National Institute of Standards and Technology (NIST)," Interagency Rep, vol. 7628, 2010
13) K. I. Sgouras, A. D. Birda and D. P. Labridis, "Cyber-attack impact on critical Smart Grid infrastructures," *IEEE PES Innovative Smart Grid Technologies conference*, Washington, DC, 2014, pp. 1-5.
14) Trong Nghia Le, Wen-Long Chin, Dang Khoa Truong and Tran Hiep Nguyen (June 29th, 2016). Advanced Metering Infrastructure Based on Smart Meters in Smart Grid, Smart Metering Technology and Services - Inspirations for Energy Utilities, Moustafa Eissa, IntechOpen.
15) S. Garlapati, H. Lin, S. Sambamoorthy, S. K. Shukla and J. Thorp, "Agent Based Supervision of Zone 3 Relays to Prevent Hidden Failure Based Tripping," *2010 First IEEE International Conference on Smart Grid Communications*, Gaithersburg, MD, 2010, pp. 256-261